# Investigation of Finite-size 2D Ising Model with a Noisy Matrix of Spin-Spin Interactions

**Boris Kryzhanovsky[1], Magomed Malsagov[1] and Iakov Karandashev[1,*]**

[1] Scientific Research Institute for System Analysis, Russian Academy of Sciences, Moscow, Russian Federation
* Correspondence: karandashev@niisi.ras.ru

**Abstract:** We analyze changes in the thermodynamic properties of a spin system when it passes from the classical two-dimensional Ising model to the spin glass model, where spin-spin interactions are random in their values and signs. Formally, the transition reduces to a gradual change in the amplitude of the multiplicative noise (distributed uniformly with a mean equal to one) superimposed over the initial Ising matrix of interacting spins. Considering the noise, we obtain analytical expressions that are valid for lattices of finite sizes. We compare our results with the results of computer simulations performed for square $N = L \times L$ lattices with linear dimensions $L = 50 \div 1000$. We find experimentally the dependencies of the critical values (the critical temperature, the internal energy, entropy and the specific heat) as well as the dependencies of the energy of the ground state and its magnetization on the amplitude of the noise. We show that when the variance of the noise reaches one, there is a jump of the ground state from the fully correlated state to an uncorrelated state and its magnetization jumps from 1 to 0. In the same time, a phase transition that is present at a lower level of the noise disappears.

---

## 1. Introduction

Calculation of the partition function is an essential of statistical physics and informatics. A few conceptual models allow exact solutions [1-6]. Among these a 2D Ising model [7], though simple, deserves special attention because of its importance for investigating critical effects. Having contributed a lot to the development of the spin glass theory, the Edwards-Anderson model [8] and Sherrington-Kirkpatrick model [9] are also worth mentioning. However, there are not many models that permit exact solutions. This is the reason why numerical methods are mostly used for tackling complex systems. Of them, two methods are most suitable for our purpose. The first is the Monte-Carlo method [10, 11]. It enables us to analyze a system and determine its critical parameters quite accurately [12-16]. The thorough consideration of the method can be found in papers [17, 18]. Unfortunately, the method needs a great deal of computations and does not allow direct calculation of the free energy. The second method uses the approach [19, 20], which has recently given rise to the fast algorithm [21, 22] that finds the free energy by computing the determinant of a matrix. The algorithm is popular because it allows the user to compute the free energy quite accurately and at the same time determine the energy and configuration of the ground state of a system.

The methods of statistical physics help researchers to understand the behavior of complex neural nets and evaluate the capacity of neural-net storage systems [23-28]. The machine learning and computer-aided image processing need fast calculations of the partition function of specific interconnect matrices [29, 30]. The realization of Hinton's ideas [31, 32] gave rise to algorithms of deep learning and image processing [33-36]. Based on the optimization of the free energy of a spin (neuron) system, the algorithms, from the formal viewpoint, comes down to the optimization of the spin correlation in neighboring layers or within a single layer of a neural net. It should be understood that the system has a phase transition because the spin correlation grows abruptly at the critical point (the correlation length becomes nearly as great as the size of the whole system). In this case the optimization of the neural network becomes temperature dependent, which makes the learning algorithm almost impracticable.

The aim of the paper is to study the properties of a finite spin system whose Hamiltonian is defined as a quadratic functional (1). The functional is often used in machine learning and image processing. Quantities $s_i = \pm 1$ may stand for either pixel class (object/background) in an image [35], or the neuron activity indication in a Bayes neural network [36]. We will use the physical notation calling quantities $s_i = \pm 1$ spins. The model under consideration has two limiting cases. The conventional 2D Ising model with regular interconnections presents the first case; the Edwards-Anderson model is the second case. The properties of our model lie somewhere in between. We introduce adjusting parameters in functional (1), which allows us to go from the 2D Ising model to Edwards-Anderson model in a smooth manner and investigate the thermodynamic characteristics of the system in the transient state.

To avoid misunderstanding, let us point out two things. First, our interest is finite systems. For this reason, there is an expected discrepancy with Onsager results obtained at $N \to \infty$. Second, we cannot use the results of the spin glass theory to the full because the finite system under consideration is ergodic: it does not have multiple phase transitions caused by frustrations and provide self-averaging [37, 38].

## 2. Essential expressions, the equation of state

Let us consider a system described by the Hamiltonian:

$$E = -\frac{1}{2N}\sum_{i \neq j}^{N} J_{ij} s_i s_j. \tag{1}$$

This system consists of $N$ Ising spins $s_i = \pm 1 (i = 1, 2, \ldots, N$, positioned at the nods of a planar grid, the nods being numbered by index $i$. Only interactions with four nearest neighbors are considered. Spin-to-spin interactions $J_{ij}$ are random and defined as

$$J_{ij} = J \cdot (1 + \varepsilon_{ij}), \tag{2}$$

where $\varepsilon_{ij}$ is a random zero-mean variable evenly distributed over the interval $\varepsilon_{ij} \in [-\eta, \eta]$. We have chosen the even distribution of $\varepsilon_{ij}$ to be able to control $J_{ij}$: when $\eta \leq 1$, all interactions are positive $(J_{ij} \geq 0)$. For the sake of simplicity, we assume that $J = 1$.

Our interest is the free energy of the system:

$$f = -\frac{1}{N}\ln Z, \tag{3}$$

where the partition function $Z = \sum_S e^{-N\beta E(S)}$ is defined as a sum over all possible configurations $S$, and $\beta = 1/kT$ is the reverse temperature. The knowledge of the free energy makes it possible to compute the basic measurable parameters of the system:

$$U = \frac{\partial f}{\partial \beta}, \quad \sigma^2 = -\frac{\partial^2 f}{\partial \beta^2}, \quad C = -\beta^2 \frac{\partial^2 f}{\partial \beta^2}, \tag{4}$$

where free energy $U = \langle E \rangle$ is the ensemble average at given $\beta$, $\sigma^2 = \langle E^2 \rangle - \langle E \rangle^2$ is the variance of energy, and $C = \beta^2 \sigma^2$ is the specific heat.

The properties of the system depend on the dimension of the system $N$ and adjusting parameter $\eta$. Unfortunately, we cannot allow for the effect of the both parameters simultaneously, so we consider the contribution of each separately.

*1) The effect of the finite grid dimension.*

Let us consider how the fact of the grid having a finite dimension affects its properties. Let us take $\eta = 0$ as the starting point. In this case the behavior of the system can be described by the expression (see reference [39]) which is true for finite systems:

$$f = -\frac{\ln 2}{2} - \ln(\cosh z) - \frac{1}{2\pi}\int_0^\pi \ln\left(1 + \sqrt{1 - \kappa^2 \cos^2\theta}\right) d\theta,$$

$$U = -\frac{1}{1+\Delta}\left\{2\tanh z + \frac{\sinh^2 z - 1}{\sinh z \cdot \cosh z}\left[\frac{2}{\pi}K_1 - 1\right]\right\}, \quad (5)$$

$$\sigma^2 = \frac{4J^2 \coth^2 z}{\pi(1+\Delta)^2} \cdot \left\{a_1(K_1 - K_2) - (1 - \tanh^2 z)\left[\frac{\pi}{2} + (2a_2 \tanh^2 z - 1)K_1\right]\right\},$$

where

$$z = \frac{2\beta J}{1+\Delta}, \quad \kappa = \frac{2\sinh z}{(1+\delta)\cosh^2 z}, \quad \Delta = \frac{5}{4L}, \quad \delta = \frac{\pi^2}{L^2},$$

$$a_1 = p(1+\delta)^2, \quad a_2 = 2p - 1, \quad p = \frac{(1 - \sinh^2 z)^2}{(1+\delta)^2 \cosh^4 z - 4\sinh^2 z}. \quad (6)$$

Here $K_1 = K_1(\kappa)$ and $K_2 = K_2(\kappa)$ are full elliptical integrals of the first and second type correspondingly:

$$K_1(\kappa) = \int_0^{\pi/2} (1 - \kappa^2 \sin^2\varphi)^{-1/2} d\varphi, \quad K_2(\kappa) = \int_0^{\pi/2} (1 - \kappa^2 \sin^2\varphi)^{1/2} d\varphi. \quad (7)$$

Expressions (5)-(7) are the well-known Onsager solution [7], which is true for $N \to \infty$, modified for the case of finite $N$. Though true for $N \gg 1$, the expressions agree well with the experimental data even at relatively small grid dimensions ($L \sim 25$. As could be expected, when $N \to \infty$, formulae (6) give $p \to 1$, $a_{1,2} \to 1$, $\Delta \to 0$, $\delta \to 0$ and expressions (5) turn into well-known ones [7].

Expressions (5) agree excellently with experimental data: the relative error is less than 0.2% at $L = 50$. With the growing $L$, the error decreases rapidly and is within the limits of experimental error at $L = 1000$ ($10^{-5}$ for $\sigma^2$). By way of comparison Figures 3, 6, and 7 gives the plots of function (5) for $L = 400$.

Expressions (5) allow the $N$-dependences of the critical values of the reverse temperature, internal energy and energy variance of the system:

$$\beta_{c0} = \beta_\infty \left(1 + \frac{1}{L}\right),$$

$$U_{c0} = -\sqrt{2}\left(1 - \frac{1}{L}\right), \quad (8)$$

$$\sigma_{c0}^2 = 2.4 \cdot (\ln L - 0.5),$$

where $\beta_\infty = \frac{1}{2}\ln(\sqrt{2}+1)$ is the critical value for $L \to \infty$ [7].

*2) The effect of noise.*

Let us consider the random character of quantities $J_{ij}$ ($\eta \neq 0$). Let $D(E)$ be the number of states of energy $E$. Then the sum of states can be presented as $Z = \sum_E D(E)e^{-N\beta E}$. Passing from summation to integration, we get (to within an insignificant constant):

$$Z \sim \int_{-\infty}^{\infty} e^{N[\Psi(E) - \beta E]} dE, \quad (9)$$

where $\Psi(E) = \ln D(E)/N$. Applying the saddle-point method to integral (9), we get $Z \sim \exp\left[Nf(\beta)\right]$, where

$$f(\beta) = \beta E - \Psi(E), \quad \frac{d\Psi(E)}{dE} = \beta. \tag{10}$$

The first expression in (10) defines the free energy, the second determines $E$ at the saddle point where the derivative of function $\Psi(E) - \beta E$ turns to zero.

The form of spectral function $\Psi(E)$ is known only for the single-dimensional Ising model. That is why we turn to the so-called n-vicinity method [28] to calculate the spectral function. The idea of the method is to divide the whole space of $2^N$ states into $N$ classes ($n$ vicinities) and approximate the energy distribution in each class by a corresponding Gaussian. In brief the approach is as follows. Let us denote the ground-state configuration as $S_0$. Let class $\Omega_n$ be a set of configurations $S_n$ that differs from $S_0$ in that they have $n$ spins directed oppositely to the spins in $S_0$. The number of configurations in the class is equal to the number of compositions of $N$ in $n$, all configurations having the same (relative) magnetization $m = N^{-1} \cdot S_m S_0^T = 1 - 2n/N$. The distribution of state energies within the n-vicinity was shown [28] to follow the normal distribution $D_n(E)$:

$$D_n(E) = \binom{N}{n} \sqrt{\frac{N}{2\pi\sigma_m^2}} \exp\left[-\frac{1}{2}N\left(\frac{E - E_m}{\sigma_m}\right)^2\right], \tag{11}$$

where

$$E_m = E_0 m^2, \quad \sigma_m^2 = 2(1 - m^2)(1 - \alpha m^2), \quad \alpha = 1 - \sigma_{h0}^2/2. \tag{12}$$

Here $E_0$ is the ground state energy, $\sigma_{h0}^2$ is the variance of ground-state local fields. In this case we have $\sigma_{h0}^2 = \sigma_\eta^2/(1 + \sigma_\eta^2)$, where $\sigma_\eta^2 = \eta^2/3$ is the variance of interconnections $J_{ij}$.

The sought-for distribution $D(E)$ is found by summing $D_n(E)$ over all $n$. Using the Stirling formula and passing from summation to integration with respect to variable $m = 1 - 2n/N$, we get for $D(E)$:

$$D(E) = \sum_{n=0}^{N} D_n(E) = \frac{N}{2\pi} \int_0^1 e^{-NF(m,E)} \frac{dm}{\sigma_n \sqrt{1 - m^2}}, \tag{13}$$

where

$$F(m, E) = -\ln N + \frac{1}{2}\left[(1 - m)\ln(1 - m) + (1 + m)\ln(1 + m) + \frac{(E - E_m)^2}{\sigma_m^2}\right]. \tag{14}$$

If we evaluate integral (13) by the saddle-point method, for the spectral function we get $\Psi(E) = -F(m, E)$, where $m$ is the solution of equation $\partial F(m, E)/\partial m = 0$. Let us combine (13)-(14) and (9)-(10). Then the free energy can be written as

$$f(\beta) = F(m, E) + \beta E, \tag{15}$$

where variables $m = m(\beta)$ and $E = E(\beta)$ are derived from the equations:

$$\ln\frac{1 + m}{1 - m} + 2\frac{E - E_m}{\sigma_m} \frac{\partial}{\partial m}\left(\frac{E - E_m}{\sigma_m}\right) = 0, \quad \frac{E - E_m}{\sigma_m^2} + \beta = 0. \tag{16}$$

It is easy to notice that set of equations (16) is solvable when $m = 0$. Correspondingly, when the values of $\beta$ are less than certain critical value $\beta_c$, (16) and (12) gives us $E_m = 0$, $\sigma_m^2 = 2$ and $E = -2\beta$, the free energy taking the form $f(\beta) = -\ln N - \beta^2$. The phase transition occurs when $\beta$ allows yet another solution to (16) at $m \neq 0$. Note that substituting the second equation from (16) into the first one allows us to eliminate variable $E$. Doing things this way and performing several transformations, we obtain the equation of state that holds only one variable $m$:

$$\frac{1}{4m}\ln\frac{1 + m}{1 - m} = \bar{\beta} - \bar{\beta}^2 (1 - m^2)\left(1 + \frac{1}{2}\sigma_\eta^2\right), \tag{17}$$

where $\bar{\beta} = \beta/r$. Here we introduced adjusting coefficient $r$ to allow for the finite grid dimension: $r = 1$ when $L \to \infty$, $r = 1.11$ giving the excellent agreement with experiments at $L \sim 400$. The critical temperature is defined as value $\beta = \beta_c$ at which there is a nontrivial solution of (17). This solution has to be found by a numerical method: when $\beta > \beta_c$, we find $m \neq 0$ that satisfies (17) and compute the corresponding value of energy $E = E_m - \beta \sigma_m^2$. Substitution of these values in (15) yields the corresponding value $f(\beta)$.

Unfortunately, the n-vicinity method has an essential fault: it is applicable only when the condition $\left(\sum J_{ij}\right)^2 / \left(N \sum J_{ij}^2\right) \geq 4 \ln 2$ holds. In our case this condition works when $(1 + \sigma_\eta^2) \cdot \ln 2 \leq 1$, that is when $\eta < 1.2$. For such relatively small values of $\eta$ formulae (15)-(17) gives $\beta_c$ and $f(\beta)$ that predict the experimental results well (see Fig.1 and Fig.2).

*3) Evaluating the spectral density.*

The algorithm we use allows us to compute function $f = f(\beta)$ and its derivatives. In turn, this allows us to investigate how energy distribution $D(E) = \exp[N\Psi(E)]$ varies with the noise amplitude. Indeed, it is easy to derive from formulae (10) the equation for the spectral function:

$$\Psi(E) = \beta E - f(\beta), \quad E = \frac{df}{d\beta} \tag{18}$$

and its derivatives

$$\frac{d\Psi}{dE} = \beta, \quad \frac{d^2\Psi}{dE^2} = \left(\frac{d^2 f}{d\beta^2}\right)^{-1}. \tag{19}$$

Note that $\Psi(E)$ is entropy up to a constant, and equations (18) are well-known Legendre transformations, which are applicable for analyzing the spectral density of finite-dimension models [40, 41]. It follows from these equations that when $\beta$ varies from $\beta = 0$ to $\beta = \infty$, $E$ changes from 0 to $E_0$, and for each value of $\beta$ we have a pair of values of $E$ and $\Psi(E)$. In so doing we determine the form of function $\Psi(E)$ and its derivatives. The plots of function $\Psi(E)$ and its derivatives presenting experimental data are given in Fig. 6-7.

Looking at Figure 7, we can see that the minimum of function $d^2\Psi/dE^2$ at point $E = 0$ changes into the maximum as the noise amplitude grows. Let us find $\eta$ at which it occurs. It can be noticed that with $E \to 0$ the entropy can be presented as the series:

$$\Psi(E) = \ln 2 - \frac{1}{2}\frac{E^2}{\sigma_J^2} + \frac{\mu_4}{4!}\frac{E^4}{\sigma_J^4}, \quad \sigma_J^2 = 2\langle J_{ij}^2 \rangle = 2(1 + \sigma_\eta^2), \tag{20}$$

where $\mu_4 = \langle E^4 \rangle / \sigma_J^4$ is the fourth cumulant, which in our case is described by the expression [28]:

$$\mu_4 = 4\left(5 - 6\sigma_\eta^2 - \frac{9}{5}\sigma_\eta^4\right) / \sigma_J^4. \tag{21}$$

From (20)-(21) it follows that in the center point of the curve ($E = 0$) quantity $d^2\Psi/dE^2$ is determined by expression:

$$\left.\frac{d^2\Psi}{dE^2}\right|_{E=0} = -\frac{1}{2(1+\sigma_\eta^2)} \tag{22}$$

and the fourth derivative $d^4\Psi/dE^4\big|_{E=0} = \mu_4/\sigma_J^4$ changes its sign at $\eta = \eta_c$, when $\mu_4 = 0$:

$$\eta_c = \left[5\left(\sqrt{2} - 1\right)\right]^{1/2}. \tag{23}$$

## 3. The experiment description

We make an intensive use of the Kasteleyn-Fisher algorithm [19, 20] here to compute the free energy of the 2D square spin system. The algorithm gives exact results because the finding of the partition function is reduced to computation of the determinant of a matrix generated in accordance with the model under consideration. The algorithm permits us to exactly calculate the free energy of a spin system for an arbitrary planar graph with arbitrary links in a polynomial time. More information about the algorithm can be found in [21]. In the paper we use the realization [22] of the algorithm that can give the same results in a shorter time. Using this algorithm, we were able to examine the behavior of free energy $f = f(\beta;\eta)$ and its derivatives for a few lattices of different dimensions $N = L \times L$. Additionally, paper [22] offers the algorithm for searching the ground state. This algorithm helped us to investigate the energy and magnetization of the ground state as functions of noise amplitude. Let us point out that the both algorithms we use are only applicable to planar lattices. It means that we considered only lattices with free boundary conditions because lattices with periodic boundary conditions do not belong to a planar graph. The length of the lattice varied from $L = 25$ to $L = 10^3$. Most of the plots present the results for $L = 400$.

The free energy is computed to 15-digit accuracy after the decimal point. Because we use the finite-difference method to compute the derivatives, the number of digits after the decimal point is about twice as less for $U(\beta)$ and four times as less for $\sigma^2(\beta)$. With large grid dimensions $(L \sim 1000$ and with $\beta > 1$ the computation error becomes too big and the plots of second derivatives start exhibiting oscillations. It is interesting to notice that introduction of little noise into grid interconnections allows us to decrease these oscillations.

## 4. Experimental results

In the experiments we calculate the free energy and its derivatives and find the ground-state configuration and energy. The accent is given to the finding of the critical point and corresponding quantities. The location of the maximum of curve $\sigma^2 = \sigma^2(\beta)$ is used to find the critical temperature. Most important experimental data are presented in Figures 1-7 and Table 1.

**Table 1.** The energy of ground state $E_0$ and its magnetization $M_0$, critical values $\beta_c$, $f_c$, $U_c$ and $\sigma_c^2$ for different noise amplitudes.

| $\eta$ | $E_0$ | $M_0$ | $\beta_c$ | $f_c$ | $U_c$ | $\sigma_c^2$ |
|---|---|---|---|---|---|---|
| 0 | -1.995 | 1 | 0.442 | -0.6931 | -1.978E-05 | 12.958 |
| 0.1 | -1.995 | 1 | 0.443 | -0.6931 | -1.986E-05 | 11.427 |
| 0.2 | -1.995 | 1 | 0.444 | -0.6932 | -0.0101 | 12.566 |
| 0.3 | -1.995 | 1 | 0.445 | -0.6932 | -0.0103 | 11.627 |
| 0.4 | -1.996 | 1 | 0.452 | -0.6933 | -0.0211 | 11.476 |
| 0.5 | -1.994 | 1 | 0.454 | -0.6934 | -0.0324 | 10.666 |
| 0.6 | -1.993 | 1 | 0.459 | -0.6936 | -0.0447 | 9.719 |
| 0.7 | -1.994 | 1 | 0.465 | -0.6939 | -0.0581 | 8.328 |
| 0.8 | -1.996 | 1 | 0.476 | -0.6946 | -0.0849 | 7.642 |
| 0.9 | -1.996 | 1 | 0.484 | -0.6957 | -0.1143 | 6.518 |
| 1.0 | -1.993 | 1 | 0.503 | -0.6979 | -0.1599 | 5.603 |
| 1.1 | -1.996 | 0.9998 | 0.515 | -0.7010 | -0.2109 | 4.656 |
| 1.2 | -1.995 | 0.9987 | 0.536 | -0.7065 | -0.2815 | 3.629 |
| 1.3 | -1.994 | 0.9943 | 0.562 | -0.7156 | -0.3747 | 2.775 |
| 1.4 | -1.996 | 0.9839 | 0.591 | -0.7327 | -0.5107 | 1.998 |
| 1.5 | -2.002 | 0.9602 | 0.623 | -0.7527 | -0.6414 | 1.380 |
| 1.6 | -2.014 | 0.9060 | - | - | - | - |
| 1.7 | -2.033 | 0.2155 | - | - | - | - |
| 1.8 | -2.065 | 0.0312 | - | - | - | - |
| 1.9 | -2.098 | 0.0241 | - | - | - | - |
| 2.0 | -2.139 | 0.0058 | - | - | - | - |

*1) The free and internal energy.*

Experimental dependencies $f = f(\beta)$ and $U = U(\beta)$ are shown in Figures 1-2. It is seen from Fig. 1 that the curves go down with $\eta$ because the ground-state energy grows. When noise is small ($\eta < 1.2$), the curves of free energy $f(\beta)$ and internal energy $U(\beta)$ almost merge (Fig. 1-2). When $\eta < 1.7$ the curves $U(\beta)$ demonstrate a cusp (Fig. 2) which corresponds to the phase transition. When $\eta \sim 1.7$, the cusp disappears and the further increase of noise changes only the asymptotic behavior of curves $f(\beta)$ and $U(\beta)$ according to (26).

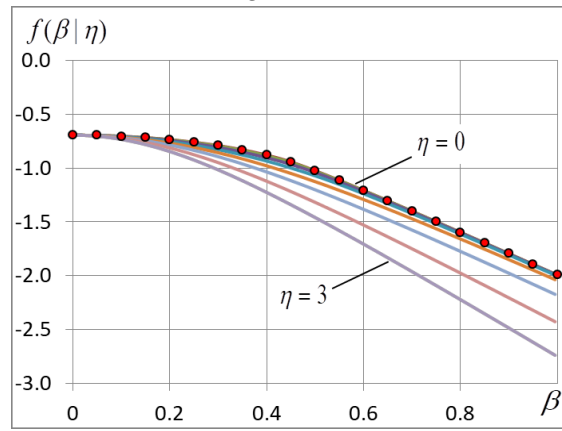

**Figure 1.** Free energy $f(\beta)$ at different noise amplitudes $\eta = 0; 0.4; 0.8; 1.2; 1.6; 2.0; 2.5; 3$. Lower curves corresponds to greater values of $\eta$. The red marks indicate the values that are found by the n-vicinity method with the aid of formulae (15)-(17) at zero noise amplitude. The grid dimension $L = 400$.

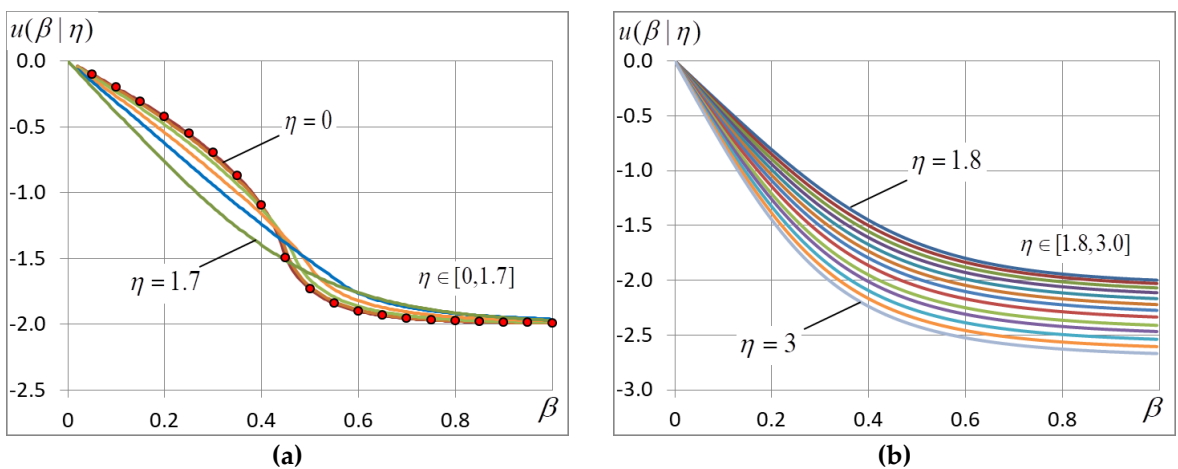

**Figure 2. (a)** Internal energy $U(\beta)$ at different noise amplitudes $\eta \in [0,1.7]$ spaced by 0.1 intervals. The red marks indicate the values that are found by the n-vicinity method with the aid of formulae (15)-(17) at zero noise amplitude. **(b)** $\eta \in [1.8, 3.0]$ spaced by 0.1 intervals, the lower curves correspond to greater $\eta$. The grid dimension $L = 400$.

*2) The energy variance.*

Curves $\sigma^2 = \sigma^2(\beta)$ are shown in Fig. 3. It should be noted that because the n-vicinity method gives a piecewise-linear approximation of the energy variance, the red marks in Fig. 3 indicates values obtained by using the generalization of Onsager solution to a finite-dimension case according to formula (5). The formula gives a perfect agreement with experimental data, yet it is applicable only in a zero-noise case.

The behavior of curves $\sigma^2 = \sigma^2(\beta)$ near point $\beta = 0$ is quite expectable for any $\eta$: when $\beta = 0$, the energy variance is equal to $\sigma_0^2$ and, according to (20), grows gradually in proportion to noise variance $\sigma_\eta^2 = \eta^2/3$. With great $\beta$ the behavior of curves $\sigma = \sigma(\beta)$ is much dependent on $\eta$. It is seen in Fig. 3 that the energy variance peaks corresponding to the phase transition are observed only at $\eta < 1.7$. The peaks become lower with growing $\eta$ and move to the right at the same time. When $\eta > 1.8$, the peaks disappear at all, only the maximum at $\beta = 0$ remains.

It is interesting that all the curves in Fig. 3 (a) have the common intersection point near $\beta \approx 0.29$. We could not find out why it is so. The intersection moves to the right slowly with the growing noise amplitude.

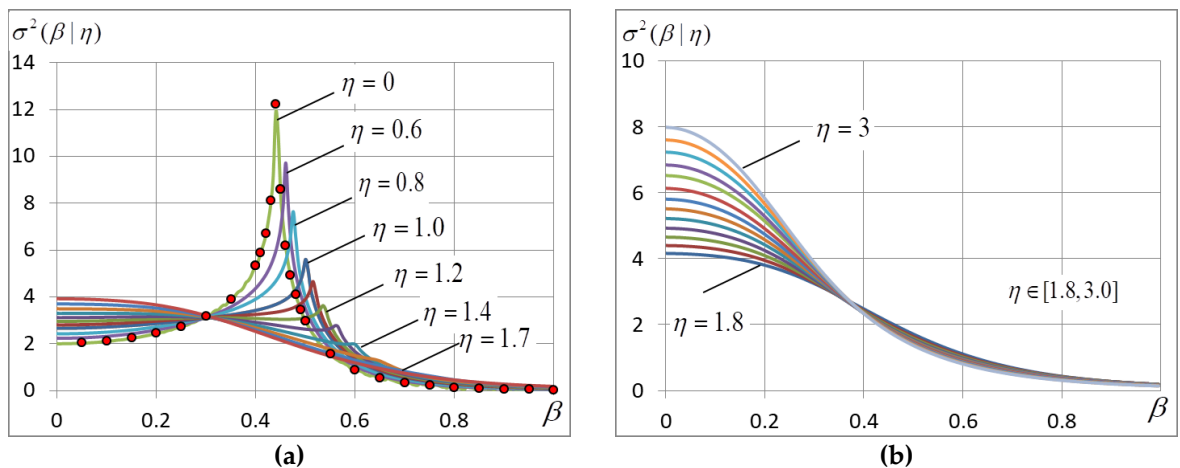

**Figure 3.** The energy variance $\sigma^2(\beta)$ at different noise amplitudes $\eta$: **(a)** $\eta \in [0,1.7]$, and **(b)** it changes by 0.1 intervals in range $\eta \in [1.8, 3.0]$. The red marks indicate values produced by formula (5). The grid dimension $L = 400$.

*3) The critical temperature.*

The critical temperature is defined by the location of the maximum of curve $\sigma = \sigma(\beta)$ or by the presence of a cusp on it. Fig. 4 shows how the variance peak location and height vary with the growing noise. Holding true only for $\eta < 1.2$, the numerical solution of the equation of state (17) gives $\beta_c$ that agrees with the experimental data perfectly. For greater $\eta$ it is possible to use the approximate expression resulting from the experiment:

$$\beta_c \simeq \beta_{c0}\left(1 - \frac{\sigma_\eta^2}{2}\right), \qquad (24)$$

where $\beta_{c0}$ is the zero-noise critical value resulted from (8). The peak height lowers linearly with the growing noise amplitude:

$$\sigma_c^2 \simeq \sigma_{c0}^2 \left(1 - \sigma_\eta\right), \qquad (25)$$

where $\sigma_{c0}^2$ is the variance at $\eta = 0$ defined in (8). It follows that if $\eta \approx \sqrt{3}$, $\sigma_c^2$ falls to zero. It means that when $\eta > \sqrt{3}$, the variance peak disappears and we can say that the critical temperature is zero.

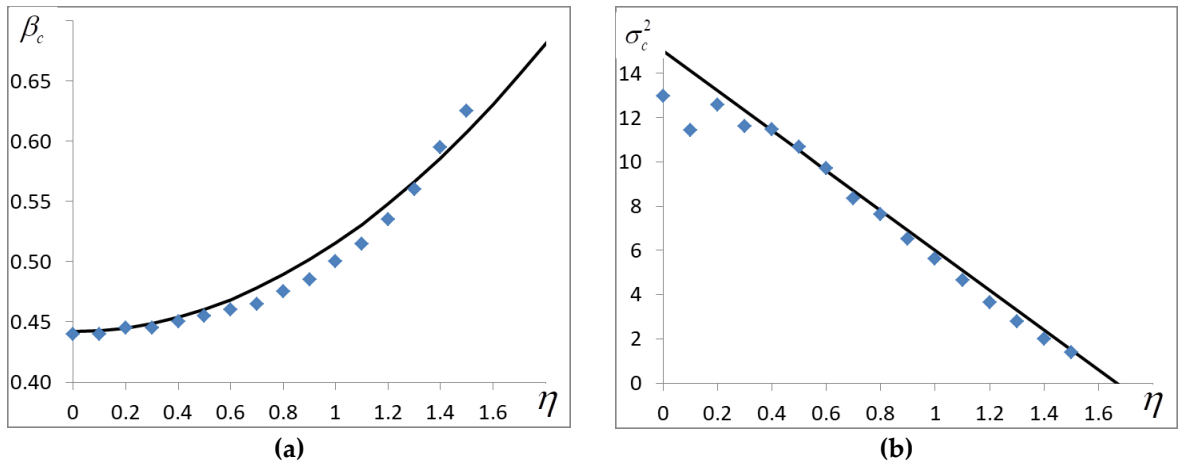

**Figure 4.** (a) The critical temperature $\beta_c$ and (b) energy variance at the critical temperature $\sigma_c^2$ as functions of noise amplitude $\eta$. The solid lines correspond to formulae (24)-(25). $L = 400$.

4) *The ground state.* The results we obtained testify that when the noise amplitude $\eta \approx 1.7$ (at $\sigma_\eta \approx 1$), the quality of the system changes. The ground state configuration experiences the most noticeable changes (see Fig. 5). Clear that with zero noise the ground state is fully correlated, i.e. all spins are the same $s_i = 1$. The situation keeps as long as all matrix elements $J_{ij} > 0$, i.e. $\eta < 1$. However, (see Fig. 5) the ground-state energy proved to remain almost the same for $\sigma_\eta$ as big as $\sigma_\eta \approx 1$. Then it starts decreasing gradually and comes to an asymptotic value [42]:

$$E_0 = -1.317\sigma_\eta, \qquad (26)$$

corresponding to the energy of the ground state in the Edwards-Anderson model. The ground-state magnetization changes stepwise from 1 to 0 when the noise deviation comes close to unit $\sigma_\eta \approx 1$.

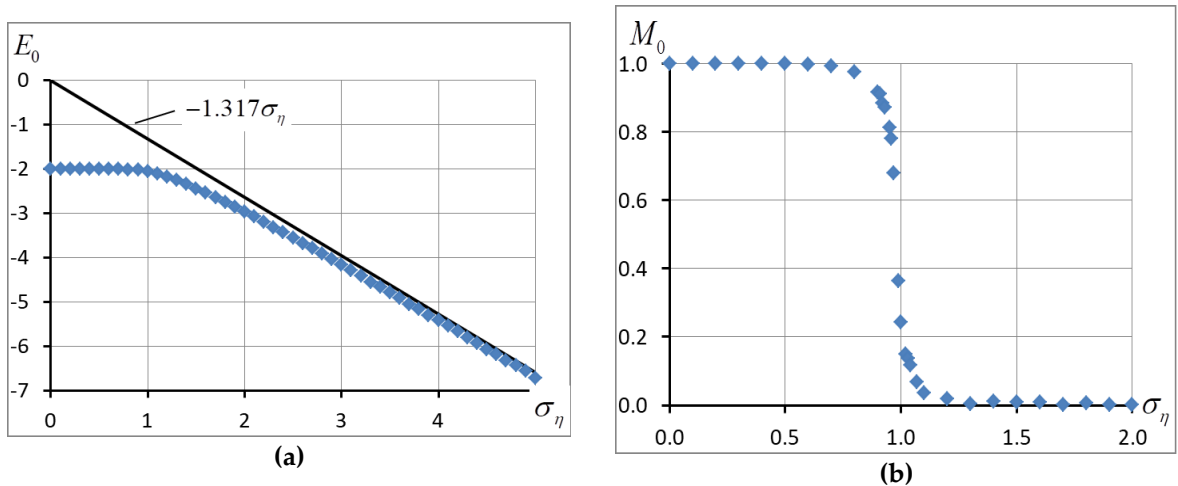

**Figure 5.** (a) Energy $E_0$ and (b) magnetization $M_0$ of the ground state of the system as a function of noise amplitude. $L = 400$.

*5) The entropy.*

The change of the ground-state configuration and energy results in a change of energy distribution density $\Psi(E)$. The curves of $\Psi(E)$ and its derivatives are shown in Fig. 6-7.

The disappearance of the phase transition is easy to notice if we look at the curve of the second derivative $d^2\Psi/dE^2$. It is seen in Fig. 7 (a) that the sink in the middle of the curve ($E = 0$) rises with growing $\eta$ and, according to (23) the minimum of $d^2\Psi/dE^2$ at $E = 0$ turns into a maximum when $\eta \approx 1.5$. The peaks at points $E = \pm U_c$ separate with growing $\eta$ ($U_c \to E_0$) and become lower like $d^2\Psi/dE^2 = -\sigma_c^{-2}$ until full disappearance at $\eta \approx 1.7$.

When $\eta > 1.7$, curve $d^2\Psi/dE^2$ has a noticeably convex shape, and the phase transition peaks disappear. Moreover, in this case function $d^2\Psi/dE^2$ is well described by the expression:

$$\frac{d^2\Psi}{dE^2} = -\frac{1}{\sigma_J^2(1-\varepsilon^2)}, \quad \varepsilon = \frac{E}{2E_0}\left(1+\frac{E^2}{E_0^2}\right). \tag{27}$$

Formula (27) gives good approximation of experimental data (accurate to 0.5% over the energy interval $0 \leq |E| \leq 0.91|E_0|$).

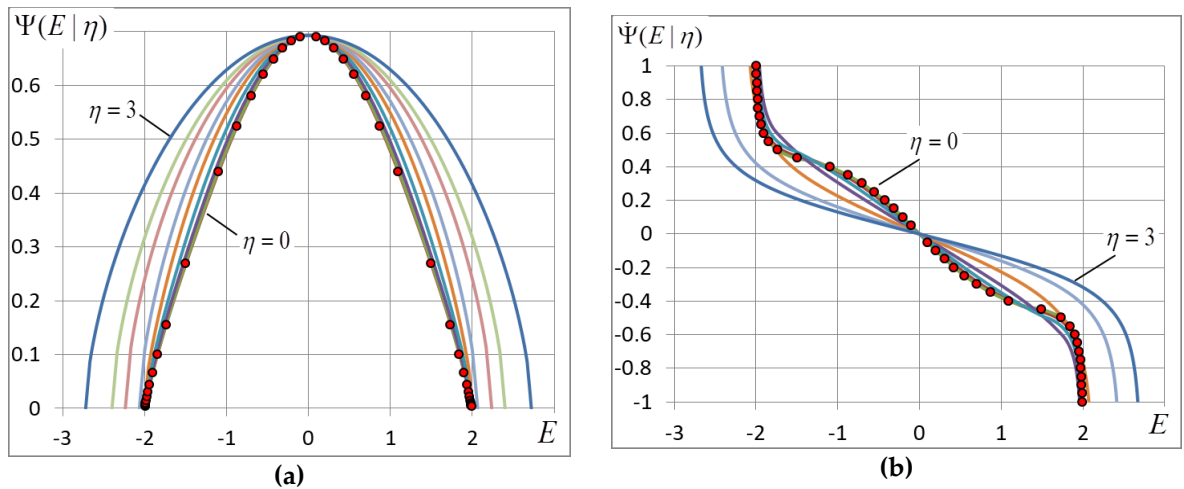

**Figure 6.** **(a)** Spectral density $\Psi(E)$ and **(b)** its first derivative for some noise amplitudes $\eta = 0; 0.3; 0.7; 1.1; 1.5; 1.8; 2.2; 2.5; 3$. The marks show the zero-noise curve. The grid dimension $L = 400$.

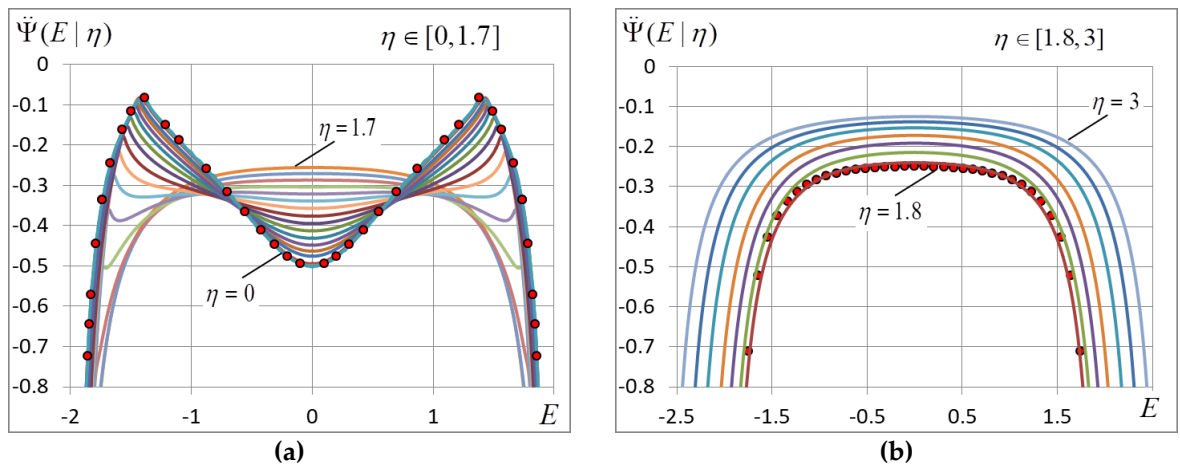

**Figure 7.** The second derivative of spectral density $\ddot{\Psi}(E)$ at **(a)** $\eta = [0,1.7]$ and **(b)** $\eta = [1.8,3]$, the reading spacing is 0.1. The marks denote the zero-noise curve **(a)** and the curve for $\eta = 1.8$ resulted from (27) **(b)**. The grid dimension $L = 400$.

## 5. Conclusions

In the paper we have considered the Ising model on a two-dimensional grid with noise-polluted interconnections. In the limiting case $N \to \infty$ such system demonstrates the following properties: with low noise the system have all characteristics of conventional Ising model, with high noise it turns into the Edwards-Anderson spin glass model. The goal of our experiments was to observe the transition between these two limiting cases in the finite-dimension system $(N \leq 10^6)$. It proved that when the noise is weak ($\sigma_\eta < 1$), the behavior of the system is much like the behavior of the conventional Ising model. We expected that with heavy noise ($\sigma_\eta \gg 1$), the behavior of the system would be like that of the Edwards-Anderson model. However, the experimental results are significantly different from the expectation. It turned out that even when the noise is relatively weak ($\sigma_\eta \sim 1$), the system undergoes considerable changes.

First, when $\sigma_\eta \sim 1$, the energy spectrum $D(E)$ changes radically (it is clearly seen in Fig. 7): the curves of $d^2\Psi/dE^2$ has a two-humped form at $\sigma_\eta < 1$, and with $\sigma_\eta > 1$ become simply convex. Moreover, the ground-state magnetization changes to zero when $\sigma_\eta > 1$. It means that when the threshold value $\eta = \sqrt{3}$ is surpassed, the ground-state configuration goes off the initial state by distance of $\frac{1}{2}N$ in the Hamming's terms. In other words, the system undergoes a zero-temperature phase transition. The transition is followed by the change of the ground-state energy from $E_0 = -2J$ to asymptotic value (26).

Second, the experimental relation between the critical temperature and noise divergence differs greatly from the well-known [8] expression $kT_c = \left(\frac{2}{9}\sum_\alpha \langle J_{i\alpha}^2 \rangle\right)^{1/2}$, which in our terms takes the form:

$$\beta_c = \frac{3}{2\sqrt{2(1+\sigma_\eta^2)}}. \tag{28}$$

We can see that the classical theory predicts that $\beta_c$ should fall with the growing deviation of noise. Moreover, expression (28) predicts finite values of $\beta_c$ for any large $\eta$. The experiment yields the opposite result: in accordance with (24) $\beta_c$ grows in proportion with $\sigma_\eta^2$. The experiment also shows that $\beta_c$ grows with $\eta$ and when $\eta \to \sqrt{3}$ it reaches its maximum $\beta_c = 0.625$, the phase transition disappears at $\eta > \sqrt{3}$ ($\sigma_\eta > 1$). It can be said conceptually that when the threshold value $\eta = \sqrt{3}$ is surpassed, the jump $T_c \to 0$ occurs.

In our opinion, the difference between the experiment and theoretical predictions is due to the finite dimension of the system. First, the finite system is ergodic and even at low temperatures does not have spontaneous magnetization, which can be tested easily with the help of Monte-Carlo algorithm. Second, the self-averaging principle used for building the theory for $N \to \infty$ is not realizable for finite $N$. Additionally, the use of terms "critical temperature" and "phase transition" is not quite correct in description of finite-dimension systems.

Finite-dimension grids are of interest in image processing and machine learning. In our paper the grid dimensions were $N = L \times L$ with $L = 25 \div 1000$. If we consider a planar grid as a model of a flat pixel image, such dimensions are very popular. The main conclusion that can be drawn from our results is that the learning algorithms based on free energy optimization are temperature insensitive in the most popular condition of $\eta \gg 1$ because there is no observable phase transition in this case.

**Acknowledgments:** The work was supported by Russian Foundation for Basic Research (RFBR Project 18-07-00750).